\documentclass[12pt]{article}
\def\be{\begin{equation}} \def\ee{\end{equation}}
\def\bi{\begin{itemize}} \def\ei{\end{itemize}}
\def\bea{\begin{eqnarray}} \def\eea{\end{eqnarray}} \def\ba{\begin{array}}
\def\ea{\end{array}} \def\ben{\begin{enumerate}} \def\een{\end{enumerate}}

\newcommand{\eqn}[1]{(\ref{#1})}

\newcommand{\plb}[3]{Phys. Lett. {\bf B#1} ({#2}) {#3}}
\newcommand{\prd}[3]{Phys. Rev. {\bf D#1} ({#2}) {#3}}
\newcommand{\hepth}[1]{{\tt hep-th/{#1}}}

\def\br{\nonumber\\}

\begin{document}
{}~
\hfill\vbox{\hbox{hep-th/yymm.nnnn} \hbox{\today}}\break

\vskip 3.5cm
\centerline{\large \bf
Special limits and  nonrelativistic solutions}
\vskip .5cm

\vspace*{.5cm}

\centerline{  
Harvendra Singh
}

\vspace*{.25cm}
\centerline{ \it  Theory Division, Saha Institute of Nuclear Physics} 
\centerline{ \it  1/AF Bidhannagar, Kolkata 700064, India}
\vspace*{.25cm}

\vspace*{.5cm}

\vskip.5cm
\centerline{E-mail: h.singh [AT] saha.ac.in }

\vskip1cm

\centerline{\bf Abstract} \bigskip

We study  special vanishing horizon limit of  `boosted' black D3-branes 
having a compact light-cone  direction. The type IIB solution 
obtained by taking such a zero temperature limit is 
found to describe a nonrelativistic  system with dynamical exponent $3$. 
We discuss about such limits in M2-branes case also.

\vfill 
\eject

\baselineskip=16.2pt


\section{Introduction}

There are mainly two type of
 non-relativistic string backgrounds, with broken Lorentzian  
symmetry, which  have been a subject 
of favorable attention currently \cite{son}-\cite{hs2}. 
The 
ones which exhibit  Schr\"odinger symmetries \cite{son,bala} are written 
as
\bea\label{sol1}
&&ds^2_{Sch}= \left( -
{\beta^2\over z^{2a}} (dx^{+})^2 -{dx^{+}dx^{-}\over z^2}
+{dx_i^2\over z^2} \right) +{dz^2\over  z^2}  
\eea 
and the
others with Lifshitz-like symmetries  \cite{Horava,kachru} are
\bea\label{sol1a}
&&ds^2_{Lif}= \left( -{\beta^2\over z^{2a}} (dt)^2
+{dx_i^2\over z^2} \right) +{dz^2\over  z^2}.  
\eea 
In both, $x^i~(i=1,...,d)$ are spatial flat coordinates, 
$z$ is the holographic scale and $a$ is called the dynamical 
exponent. 
These  geometries have been  claimed 
  to be describing  interesting scaling phenomena near
quantum critical points in the dual  CFTs \cite{son,bala}.
The Schr\"odinger  AdS spacetimes can be embedded in string theory
as  was shown in  \cite{herzogrev,malda}. These spaces can even be 
obtained in the massive type IIA 
string theory \cite{hs}. 
More recently, 
the Lifshitz-like spaces have been embedded in string theory as 
 demonstrated by \cite{bala3} and  generalised by
 \cite{donos10}. 
This implies that a wide class of
non-relativistic solutions can be found in string theory and the hope 
is that
these could potentially describe  interesting scaling 
phenomena in dual field theories on the boundary.
For  study of finite
temperature properties like phase transitions, entropy etc. 
one includes black holes in the AdS backgrounds.

In this work we shall discuss a new type of Galilean AdS geometry which
has one of the lightcone 
direction, namely $x^{+}$, being null while the other, $x^{-}$, being
compact
\bea\label{sol1b}
&&ds^2= \left( -{dx^{+}dx^{-}\over z^2}+
{\beta^2 z^{2}\over 4 } (dx^{-})^2
+{dx_1^2 +dx_2^2\over z^2}\right) +{dz^2\over  z^2}  
\eea 
It is a unique  non-relativistic solution  directly 
obtainable by taking  `zero' 
temperature (condensation) limits of a boosted black D3-brane AdS 
geometry. 
Indeed we find that the corresponding thermodynamic 
quantities in the dual theory undergo a kind of condensation where
 \be
 T\sim 0,~~~\mu_{_N}\sim 0,~~~s\sim0,
~~~~\rho \sim {T^4\over \mu_{_N}^3}={\rm fixed}.
\ee
Our study shows that such  thermodynamic limits do exist which 
directly lead us to such non-relativistic Lifshitz systems at zero 
temperature.
The zero temperature non-relativistic theory  dual to
the Galilean geometry \eqn{sol1b}  however  
exhibits a scaling symmetry with dynamical exponent  $a=3$.  

The paper is organised as follows. In  section-II we describe the 
boosted black D3-brane AdS solution. We then go over to discuss a 
special  $r_0\to 0,~\lambda\to\infty$ limit in which the black hole 
horizon is allowed to shrink 
while the boost is simultaneously taken to be very large. The geometry 
thus obtained 
describes a  `zero' temperature nonrelativistic dynamics in a 
2-dimensional 
(planar) quantum mechanical system. 
In  section-III we discuss a type IIA Lifshitz solution obtained under 
T-duality. The section-IV consists
of taking  similar limits of the boosted black M2-branes. The dual 
 theory 
represents a nonrelativistic phenomenon with fractional scaling power in 
one space dimension. The 
conclusions are 
given in section-V.

 \section{ Vanishing horizon limits}

\subsection{DLCQ and $AdS_5$ space }
We are interested in studying the DLCQ of the  $AdS_5$ space as 
described in \cite{malda}.
We start with the boosted version of black D3-branes  \cite{malda}
where the near horizon geometry is
\bea\label{sol2}
&&ds^2_{D3}=r^2\left( -{1+f\over 2}dx^{+}dx^{-}+{1-f\over 
4}[{(dx^{+})^2\over 
\lambda^2} +\lambda^2 (dx^{-})^2]
+dx_1^2 +dx_2^2\right) 
\br && ~~~~~~~~~~ 
+{dr^2\over f r^2}  + d\Omega_5^2 \, ,\br
&& F_5=4(1+\ast)
Vol(S^5)
\eea 
where $d\Omega_5^2$ represents the 
line element of a unit  five-sphere while $Vol(S^5)$ is its volume element. 
The overall $AdS_5$ radius has been set to unity.
Here $f(r)=1-r_0^4/r^4$ with 
$r=r_0$ being  the horizon location and the boundary is at $r\to\infty$. 
The boundary conformal field theory has finite temperature.  
These black 3-branes have large (finite) momentum along $x^{-}$, 
which is  compact, $x^{-}\sim x^{-}+2\pi r^-$. 
If we try to set $r_0=0$  it would make $x^-$ circle  null for which we 
should be careful. 
It usually is not a problem when
$x^{-}$ is noncompact, because then setting $r_0=0$  simply is the
extremal 
limit which gives 
 ordinary AdS spacetime whose dual  is a 
super-Yang-Mills theory at large 't Hooft coupling, 
$g_{_{YM}}^2 N'$. ( $N'$ is the order of the gauge group 
and also the number of D3-branes which give rise to the AdS geometry.)
Note that the size of $x^-$ 
circle anyway vanishes as we go near the boundary at $r\to\infty$, but 
nevertheless it stays finite in the region inside the bulk. 
Physically, the boost (scale) parameter
$\lambda$ controls the size of $x^{-}$ circle. 

\subsection{The simultaneous $r_0\to 0,~\lambda\to\infty$ limit}
As we learnt, taking $r_0=0$ is not possible  in 
\eqn{sol2} when $x^{-}$ is compact. But, 
since $r_0^4\lambda^2$ effectively controls the size of the 
circle  we would instead consider
a simultaneous limit in which the size of 
horizon is allowed to shrink  while boost is taken to infinity such that
\be\label{sol21}
r_0\to 0,~~~~ \lambda\to\infty, ~~~~r_0^4\lambda^2=\beta^2= {\rm fixed}.
\ee
 In which case we can have
\bea\label{sol22}
&&(1+f)\to 2+O(r_0^4),~~~~  {1-f\over\lambda^2}\to  O({r_0^4\over\lambda^2}) \br
&& {(1-f)\lambda^2}\to  {\beta^2\over r^4}
\eea
and the solution \eqn{sol2} reduces to
\bea\label{sol23}
&&ds^2_{D3}\simeq r^2\left( -dx^{+}dx^{-}+{\beta^2\over4 r^4} (dx^{-})^2
+dx_1^2 +dx_2^2\right) +{dr^2\over  r^2}  + d\Omega_5^2  \br
&& F_5=4(1+*)Vol(S^5) 
\eea 
which is a complete solution of type IIB string theory. The 
coordinate $r$  can be 
identified with $1/z$ in \eqn{sol1b}.
Actually spacetime \eqn{sol23} 
is  an $AdS_5$ geometry but it is Galilean one
and with $x^{+}$ being  null. 
However it should not  to be worried since $x^{-}$ is a circle
and we can always rewrite Eq.\eqn{sol23} 
in a diagonal basis as
\bea\label{sol24}
ds^2_{D3}&=& 
r^2\left( -{r^4\over \beta^2}(dx^{+})^2+
{\beta^2\over 4 r^4} 
(dx^{-}-{2r^4\over\beta^2}dx^{+})^2
+dx_1^2 +dx_2^2\right) +{dr^2\over  r^2}  + d\Omega_5^2  \br
&=&\left( -{r^6\over \beta^2}(dx^{+})^2
+r^2(dx_1^2 +dx_2^2) +{dr^2\over  r^2}\right) +
{\beta^2\over 4 r^2} 
(dx^{-}-{2r^4\over\beta^2}dx^{+})^2  + d\Omega_5^2 \br 
\eea 
From above eq.\eqn{sol24} it is obvious that
 the Galilean geometry \eqn{sol23} indeed represents a  well defined
system of  Kaluza-Klein particles if we ever dimensionally reduce over 
$x^{-}$ and go to lower dimensions.
That is we are simply dealing with bosonic KK excitations in one less 
dimensions. 
However, unlike \eqn{sol2} the new geometry \eqn{sol23} 
 is inherently nonrelativistic. The line element 
$\left( -{r^6\over \beta^2}(dx^{+})^2
+r^2(dx_1^2 +dx_2^2) +{dr^2\over  r^2}\right)$ in eq.\eqn{sol24}
is precisely the Lifshitz
four-universe.
 Hence we find that by performing the
 special  vanishing horizon limit \eqn{sol21} on the `boosted' black 
D3-branes allows us to exclusively  zoom onto a KK system in a Lifshitz 
universe. 
There is a non-relativistic scale (dilatation) invariance
\be
r\to (1/\xi) r , ~~~~
x^{-}\to \xi^{2-a}  x^{-},~~~~
x^{+}\to \xi^a x^{+},~~~x_{1,2}\to \xi x_{1,2}
\ee
with  dynamical exponent $a=3$. There are also
  invariances under constant shifts (translations) like
\be
x^+ \to x^+ + b^{+},~~~~
x^- \to x^- + b^{-},~~~~x^i\to x^i +b^i
\ee
 as well as   rotations in the $x^1-x^2$ plane
\be
x^i\to  \omega^i_j ~x^j.
\ee
 However, \eqn{sol23} does not have  any explicit invariance under
the  Galilean boosts 
\be\label{boos}
x^+\to x^+,~~~x^-\to x^--\vec v.\vec x +{v^2\over2}x^+,~~~\vec 
x\to~\vec x-\vec v x^+.
\ee
Same is the case with the special conformal transformations.

Thus our solution \eqn{sol23} represents a (Lifshitz) geometry with a 
broken Lorentzian 
symmetry in which the time scales with  a
dynamical exponent   $a=3$. 
Note, however, no extra matter fields are present in
the background \eqn{sol23} except the self-dual 5-form flux. This is 
unlike many other Schr\"odinger geometries, e.g. in 
\cite{herzogrev,malda}, 
obtained via T-s-T mechanism, where $B_{\mu\nu}$ field 
contributes as `dust' to the energy-momentum tensor. But we do have 
 Kaluza-Klein excitations present in \eqn{sol23}. 
We 
 have checked that  the  background \eqn{sol23}
preserves at least 8  Poincar\'e supersymmetries.

\subsection{Limits of thermodynamic quantities}
Having obtained the  nonrelativistic geometry in \eqn{sol23} 
we shall 
now study the effect of the $r_0\to 0,~\lambda\to\infty$ limits on the 
thermodynamic 
quantities. As we know that the  hot boundary CFT involves a DLCQ 
description  \cite{malda}. 
The isometry along $x^{-}$  implies that there is a conserved
 momentum (charge) $P_{-}$ in the theory which is quantized in units of 
${1\over r^-}$. The number (charge) density  depends upon 
the choice of two parameters, namely $\lambda$ and $r_0$.   
We want to  know what happens to the
 number density, energy density
$(-P_{+})$ and other thermodynamical quantities; like temperature $(T)$, 
entropy $(S)$ 
and chemical potential $(\mu_{_N})$ relevant
for the black-hole solution \eqn{sol2} as we consider  special  limits 
\eqn{sol21}. 
We can find  these estimates simply
by using the thermodynamic expressions given in \cite{malda}
\bea\label{tq}
&& \rho={N\over v_2}={r^{-}(-P_{-})\over v_2}
={L^3\over G_N^5}{(r^{-})^2\lambda^2 
r_0^4\over 8} \sim {L^3\over G_N^5}{(r^{-})^2\beta^2\over 8}    \br
&& {\cal E}={H\over v_2}={(-P_{+})\over v_2}={L^3\over G_N^5}{r^{-} 
r_0^4\over 16}\sim O({r_0^4}) \sim 0\br
&&s= {S\over v_2}={L^3\over4 G_N^5}(2\pi r^{-}){\lambda 
r_0^3\over 2} \sim O(r_0)\sim 0 \br
&& { T}={r_0\over \pi \lambda}\sim O(r_0^3)\sim 0,
~~~\mu_{_N}={1\over r^{-}\lambda^2}\sim O(r_0^4)\sim 0 
  \eea
where $L(=1)$ is the $AdS_5$ radius, 
$G_N^5$ is 5-dimensional Newton's constant
and $v_2$ is the volume of 
$x_1-x_2$ plane. We see that under our special 
  limits \eqn{sol21}, the temperature of  boundary $(2+1)$ dimensional 
 theory effectively  vanishes
so also the entropy and the chemical potential.
Curiously though, we  find that  the  number 
density remains fixed while energy density altogether 
vanishes. Although  these quantities are vanishing,  
worth noticing is their unique scaling behaviour as powers of 
vanishing horizon radius $r_0$
\be\label{tq1}
T\sim r_0^3, ~~~{\cal E}\sim r_0^4, ~~~~\mu_{_N}\sim r_0^4, ~~~~s\sim r_0
\ee
The temperature vanishes as $r_0^3$
which is an indication of the fact that system becomes nonrelativistic 
with
 dynamical exponent 3 as it undergoes condensation. 
 \footnote{In the DLCQ set-up \cite{malda}, 
the lightcone Hamiltonian in a given
 momentum sector $(P_-)$ reads as 
$H=-P_{+}\sim{\vec p^2\over (-4P_{-})}+\cdots$. 
Here conserved momentum $(-P_-)$ plays the role of Galilean mass $M$.}  
Thus it would be appropriate to 
 call the `vanishing' horizon 
 limits \eqn{sol21} as  `zero' temperature (or condensation) limits 
 \be
T\to 0,~~~\mu_{_N}\to 0,~~~~\rho \sim {T^4\over  \mu_{_N}^3}={\rm fixed}.
\ee
in the  DLCQ  theory. 
 Moreover the scaling 
behaviour in \eqn{tq1} summarises how this condensation would be 
achievable such that at the end point
we have a zero temperature nonrelativistic DLCQ theory with dynamical 
exponent 
$a=3$. In this sense this zero temperature limit of the DLCQ is very 
special. To recall the thermal 
 $D=4$  super Yang-Mills theory,  dual to nonextremal
D3-branes, has a zero temperature (extremal) 
limit $r_0\to0$ under which $T\to r_0,~s\to r_0^3$. The end point of this 
condensation leads us to ordinary ${\cal N}=4$ super Yang-Mills theory 
which is 
a relativistic theory. 

If we wish,
from eqs.\eqn{tq} we can also write down the standard thermodynamical 
expressions for free energy density and the entropy density 
\bea
&& F\sim -{\cal E}\sim -{T^4\over\mu_{_N}^2}\sim 0,~~~~~
s \sim {T^3\over  \mu_{_N}^2}\sim 0
\eea 
which are vanishing  under our zero temperature limit.
So this describes a condensate in DLCQ and 
there is only one relevant parameter and that 
is overall number density ($\rho$)  in a given 
condensate which depends on $\beta$. 
\footnote{ An interesting study 
on condensation in Galilean ideal gases can be found in \cite{barbon}.}  
We have the following comments to offer.
\bi
\item
The thermodynamic quantities in \eqn{tq} usually 
satisfy the first law of thermodynamics 
\be
\delta H(T,\mu_{_N})=T \delta S-\mu_{_N}\delta N
\ee
Since both $(T,~\mu_{_N})\to 0$ we  see that energy fluctuations 
$\delta H$ are very much suppressed. This provides stability to the 
ground state under thermal fluctuations. 

\item
Note,  our classical 
geometry \eqn{sol23} cannot be trusted near the AdS boundary
 as the 
physical size of  $x^{-}$ circle 
\be
{R^{-}_{phys}\over l_s}={L\over l_s}{r_0\over r}{r^{-}\lambda 
r_0\over 2}\sim \beta{r^{-}\over r} 
\ee
vanishes near $r\to\infty$.  
It will  then be appropriate to go over to 
type IIA string picture where the T-dualised circle 
could have a finite radius. We shall discuss more about it next.

\item
We have a well controlled zero temperature limits both within the 
bulk and in the boundary theory. Thus we can safely conclude that 
the zero temperature  
geometry \eqn{sol23} is genuine and its dual boundary theory
 will exhibit a Lifshitz symmetry with 
dynamical exponent $a=3$ at least in the IR region. In the UV region we 
will have to look for an appropriate M-theory picture as we discuss it 
next.
 \ei

\section{The type IIA solution}
We noticed that the size of the compact null direction in type IIB 
solution \eqn{sol23} 
becomes sub-stringy as we approach the boundary region. So it would be 
appropriate to make T-duality transformation along the $x^{-}$ circle in
 \eqn{sol23} where its radius gets inverted. We  write down  
corresponding type IIA background as
\bea\label{sol3}
ds^2_{IIa}&= & \left( -{r^6\over\beta^2}(dx^{+})^2+
r^2({4\over \beta^2} (d\tilde x^{-})^2
+dx_1^2 +dx_2^2) +{dr^2\over r^2} \right)   + d\Omega_5^2  \br
&=&
\left( -{1\over\beta^2 z^6}(dx^{+})^2+
{ 
dx_1^2 +dx_2^2+dz^2\over z^2}  \right) +{4\over\beta^2 z^2}(d 
\tilde x^{-})^2
  + d\Omega_5^2 , \br
e^{2\phi_a}&=&g_s^2({4\over\beta^2 z^2}),~~~~ 
A_{+12}\sim-{1\over  z^4},~~~
B_{-+}^{NS}\sim {2\over\beta^2z^4}
\eea  
The circle $\tilde x^-$ is the T-dual circle in type IIA.
The above solution has a geometry in which
a 4-dimensional  Lifshitz spacetime has a circle $\tilde x^{-}$ foliated
over it.
However the 
dilaton and the NS $B$-field are also present.
There is a new Lifshitz type scaling invariance
\be\label{lif1}
z\to \xi z , ~~~~
x^{+}\to \xi^{2} x^{+},~~~x_{1,2}\to \xi x_{1,2}
\ee
provided that  $\beta\to \xi^{-1} \beta$. Note that $\tilde x^{-}$
does not scale, this scaling is broken in type IIA solution.
This is the effect of T-duality along $\tilde x^-$.
But since $\beta$ is a free parameter effectively controlling the size 
of transverse
$\tilde x^{-}$ circle, from type IIA point of view
the Lifshitz scaling \eqn{lif1} would take us
 from one type IIA picture to an  equivalent type IIA picture. 

We see from type IIA solution \eqn{sol3} that the string 
coupling   runs 
and it becomes large in UV. As we go close to the boundary 
the type IIA description 
will break down and one would need to go over to an appropriate 
M-theory picture. 
The M-theory geometry  obtained by uplifting the 
solution \eqn{sol3} to eleven dimensions is
\bea\label{ph1}
&&ds_{M}^2 
= ({\beta^2 z^2\over 4g_s^2})^{1\over3}\left( -{(dx^{+})^2\over \beta^2
z^6}+{dx_1^2+dx_2^2 +dx_{1\!1}^2+(d\tilde x^{-})^2+ 
dz^2 \over z^2}+d\Omega_5^2\right) \br
&&~~~~~~\simeq
 ({\beta^2\over 4 z^4})^{1/3}\left( [-{1\over 
\beta^2 z^4}(dx^{+})^2+dx_1^2+dx_2^2]
+dx_{1\!1}^2+(d\tilde x^{-})^2+ 
(dz^2+z^2d\Omega_5^2)\right)\br
&& C_{+12}\sim {1\over  z^4},~~~~~ C_{+-1\!1}\sim{1\over\beta^2z^4}
\eea
where circular coordinate $x_{1\!1}$ 
has been appropriately scaled. This describes a collection 
of $N'$ M2-branes stretched along $x^1-x^2$ plane and living
over transverse 8-dimensional space $S^1\times \tilde S^1\times R^6$.
We observe that the solution \eqn{ph1} still has a singularity near 
$z=0$  as $(F_{(4)})^2$ would blow up there
and it would be  inevitable to include quantum corrections.

\section{ Limits for black M2-branes}
Similar to the zero-temperature limits of the DLCQ in $AdS_5$ space, 
we can also study the similar limits for DLCQ of M2-brane theory.  
We start with the boosted version of black M2-branes (with a compact 
lightcone circle) 
where the near horizon geometry can be written as
\bea\label{sol41}
&&ds^2_{M2}=r^4\left( -{1+f\over 2}dx^{+}dx^{-}+{1-f\over 
4}[{(dx^{+})^2\over 
\lambda^2} +\lambda^2 (dx^{-})^2]
+dy^2\right) +{dr^2\over f r^2}  + d\Omega_7^2  ,\br
&& F_4=6Vol(AdS_4)
\eea 
where $d\Omega_7^2$ is the 
line element of a unit  seven-sphere and $\lambda$ is the boost 
parameter. 
Here $f(r)=1-{r_0^6\over r^6}$ and 
$r=r_0$ is the horizon location for the black M2-branes.  
Taking the vanishing horizon limits analogous to the D3-brane 
case
\be\label{sol42}
r_0\to 0,~~~~ \lambda\to\infty, ~~~~r_0^6\lambda^2=\beta^2={\rm fixed},
\ee
 the geometry \eqn{sol41} becomes
\bea\label{sol43}
ds^2_{M2}
&\simeq & r^4\left( -dx^{+}dx^{-}+{\beta^2\over4 r^6} (dx^{-})^2
+dy^2\right) +{dr^2\over  r^2}  + d\Omega_7^2  \br 
&=& {\tilde r^2}\left( -dx^{+}dx^{-}+{\beta^2\over4 \tilde r^3} 
(dx^{-})^2
+dy^2\right) +{d\tilde r^2\over 4 \tilde r^2}  + d\Omega_7^2 ,
\eea 
while $F_4$ flux remains unchanged. 
The scale invariance could be read as
\be
\tilde r\to (1/\xi) \tilde r , ~~~~
x^{-}\to (1/\sqrt{\xi})  x^{-},~~~~
x^{+}\to \xi^{5/2} x^{+},~~~y\to \xi^{} y .
\ee
There is also shift invariances along $x^\pm$ and $y$ but no Galilean 
boost invariances 
as such. 
The geometry \eqn{sol43} would then describe a zero 
temperature 
quantum phenomenon along a wire in the Galilean  theory.   
It is interesting to notice that in this M-theory example the time scales
with a `fractional' dynamical exponent, which is $5/2$.

Once again in the UV region $\tilde r \gg \beta^2/4$ the size of the spatial 
circle becomes very small in 11-dimensional Planck length units and we 
cannot trust
our classical solution there. It would require to study quantum 
corrections to the geometry or to add appropriate boundary corrections so 
as to 
make it UV picture complete.

\section{Conclusion}
We have studied special `vanishing' horizon limits
of  the `boosted' black D3-branes 
having a compact  lightcone direction. The limits are taken in such 
a way that the compact direction does not become null so that we can have 
DLCQ description of the CFT.
The resultant zero temperature solution is 
found to describe a non-relativistic system with Lifshitz symmetry having 
a dynamical exponent $a=3$.
The new Lifshitz
geometry however cannot describe the UV regime  as the size of 
lightcone circle becomes sub-stringy as we approach the AdS boundary. On 
the other hand the string coupling 
grows very large in the UV region in a corresponding T-dual type IIA 
picture, 
which indicates that it will
 require us to look for a M-theoretic interpretations. It will therefore 
be interesting to find a resolution to this UV problem.
It would also be worthwhile to explore the zero-temperature limits like 
in our work for 
other 
`boosted' black D$p$-brane solutions and study the  resulting Lifshitz 
solutions thus obtained. 
We have discussed an example of  the black M2-branes here. The Galilean 
geometry obtained shows that the scaling symmetry has fractional dynamical 
exponent.




\end{document}